\input harvmac
     
\Title{\vbox{\rightline{EFI-2001-08}\rightline{MIT-CTP-3098}
\rightline{hep-th/0103102}}}
{\vbox{\centerline{Boundary Superstring Field Theory}}}
\medskip

\centerline{Vasilis Niarchos\footnote{$^a$}{vniarcho@theory.uchicago.edu} and
Nikolaos Prezas\footnote{$^b$}{prezas@mit.edu}}
\bigskip
\centerline{$^a$\sl Enrico Fermi Inst. and Dept. of Physics}
\centerline{\sl University of Chicago}
\centerline{\sl 5640 S. Ellis Ave., Chicago, IL 60637, USA}
\smallskip
\centerline{$^b$\sl Center for Theoretical Physics}
\centerline{\sl Massachusetts Institute of Technology}
\centerline{\sl 77 Mass. Ave., Cambridge, MA 02139, USA}
 
\smallskip

\vglue .3cm
\bigskip

\noindent

Using the Batalin-Vilkovisky formalism we provide 
a detailed analysis of the $NS$ sector of boundary 
superstring field theory. We construct explicitly the relevant
BV structure and derive the master action. Furthermore, we
show that this action is exactly equal to the superdisk 
worldsheet partition function as was recently conjectured.

\Date{3/2001}

%%%%%%%%%%%%%%%%%%%%%%%%%%%%%%%%%%%%%%%%%%%%%%%%%%%%%%%%%%%
%    Defs
%
%

\def\frac#1#2{{#1\over#2}}

\def\inbar{\,\vrule height1.5ex width.4pt depth0pt}
\def\IC{\relax\hbox{$\inbar\kern-.3em{\rm C}$}}
\def\IR{\relax{\rm I\kern-.18em R}}
\def\IP{\relax{\rm I\kern-.18em P}}

%
%%%%%%%%%%%%%%%%%%%%%%%%%%%%%%%%%%%%%%%
%

%
\catcode`\@=11
\def\slash#1{\mathord{\mathpalette\c@ncel{#1}}}
\overfullrule=0pt
\catcode`\@=12

%%%%%%%%%%%%%%%%%%%%%%%%%%%%%%%%%%%%%%%

%

%%%%%%%%%%%%%%%%%%%%%%%%%%%%%%%%%%%%%%%%%%%%%%%%%%%%
%%%%%%%%%%%%%%%%%%%%%%%%%%%%%%%%%%%%%%%%%%%%%%%%%%%%
%      References
%
\lref\witone{E.~Witten, ``On Background Independent 
Open-String Field Theory,''
Phys.\ Rev.\   {\bf  D47} (1992) 5467; hep-th/9208027.}
\lref\wittwo{E.~Witten, ``Some Computations in Background Independent 
Open-String Field Theory,'' Phys.\ Rev.\   {\bf  D47} (1993) 3405;
hep-th/9210065.}
\lref\shataone{S.~Shatashvili, ``Comment on the Background Independent 
Open String Theory,'' Phys.\ Let.\   {\bf  B311} (1993) 83;
hep-th/9303143.}
\lref\shatatwo{S.~Shatashvili, ``On the Problems with Background Independence in String Theory,'' hep-th/9311177.}
\lref\kutone{D.~Kutasov, M.~Marino, G.~Moore, ``Some Exact Results on 
Tachyon Condensation in String Field Theory,''
JHEP  {\bf 0010} (2000) 045; hep-th/0009148.}
\lref\kuttwo{D.~Kutasov, M.~Marino, G.~Moore, ``Remarks on Tachyon 
Condensation in Superstring Field Theory,'' hep-th/0010108.}
\lref\andre{O.~Adreev, ``Some Computations of Partition 
Functions and Tachyon Potentials in Background Independent,'' hep-th/0010218.}
\lref\gs{M.~Greene, N.~Seiberg,``Contact Interactions in Superstring Theory,''
Nucl.\ Phys.\   {\bf  B299}, (1988) 559.}
\lref\sw{N.~Seiberg, E.~Witten, ``String Theory and Noncommutative Geometry,''
JHEP  {\bf 9909} (1999) 032; hep-th/9908142.}
\lref\taka{T.~Takayanagi, S.~Terashima, T.~Uesugi, ``Brane-Antibrane Action from
Boundary String Field Theory,'' hep-th/0012210.}
\lref\larsen{P.~Kraus, F.~Larsen, ``Boundary String Field Theory of the
DDbar System,'' hep-th/0012198.}
\lref\tseytone{O.~Andreev, A.~Tseytlin, ``Partition Function Representation
for the Open Supestring Action,'' Nucl.\ Phys.\ 
{\bf  B311} (1988) 205.}
\lref\tseyttwo{A.~Tseytlin, ``Sigma Model Approach to String Theory,'' Int.\ Jour.\
Mod.\ Phys. \ {\bf  A4} (1989) 1257.}
\lref\tseytthree{A.~Tseytlin, ``Renormalization Group and String Loops,'' Int.\ Jour.\
Mod.\ Phys. \ {\bf  A5} (1990) 589.}
\lref\affone{I.~Affleck, W.~Ludwig, ``Universal Noninteger ``Ground State
Degeneracy'' in Critical Quantum Systems,'' Phys.\ Rev.\ Lett.\  {\bf 67}
(1991) 161.}
\lref\afftwo{I.~Affleck, W.~Ludwig, ``Exact Conformal Field Theory 
Results on the Multichannel Kondo Effect: Single Fermion Green's Function, 
Self-energy, and Resistivity,'' Phys.\ Rev.\   {\bf B48} (1993) 7297.}
\lref\fms{D.~Friedan, E.~Martinec, S.~Shenker, ``Conformal Invariance,
Supersymmetry, and String Theory,'' Nucl.\ Phys.\ {\bf B27} (1986) 93.}
\lref\belopo{A.~Belopolsky, ``Picture Changing Operators in Supergeometry and 
Superstring Theory,'' hep-th/9706033.}
\lref\liwit{K.~Li, E.~Witten, ``Role of Short Distance Behavior in Off-Shell Open-String
Field Theory,'' Phys.\ Rev.\ {\bf D48} (1993) 853; hep-th/9303067.}
\lref\gomis{J.~Gomis, J.~Paris, S.~Samuel, ``Antibracket, Antifields and Gauge Theory 
Quantization,'' hep-th/9412228.}
\lref\zwione{B.~Zwiebach, ``Closed String Field Theory: Quantum Action 
and the Batalin-Vilkovisky 
Master Equation,'' Nucl.\ Phys.\ {\bf B390} (1993) 33; hep-th/9206084.}
\lref\zwitwo{B.~Zwiebach, ``Closed String Field Theory: An Introduction,'' hep-th/9305026.}
\lref\thorn{C.~Thorn, ``String Field Theory,'' Phys.\ Rep.\ {\bf 174} (1989) 1.}
\lref\bochi{M.~Bochicchio, ``Gauge Fixing for the Field Theory of the Bosonic String,''
Phys.\ Lett.\ {\bf B193} (1987) 31.}
\lref\henneaux{M.~Henneaux, C.~Teitelboim, ``Quantization of Gauge Systems,'' Princeton
University Press (1992).}
\lref\polch{J.~Polchinski, ``String Theory,'' Vol. 2, 
Cambridge University Press (1998).}
\lref\bataone{I.~Batalin, G.~Vilkovisky, ``Quantization of Gauge Systems
with Linearly Dependent Generators,'' Phys.\ Rev.\ {\bf D28} (1983) 2567.}
\lref\batatwo{I.~Batalin, G.~Vilkovisky, ``Existence Theorem for Gauge
Algebras,'' J.\ Math.\ Phys.\ {\bf 26} (1985) 172.}
\lref\sen{A.~Sen, ``Descent Relations Among Bosonic D-branes,'' Int.\ J.\ Mod.\ Phys.\
{\bf A14} (1999) 4061.}
\lref\witipco{E.~Witten, ``Interacting Field Theory of Open Superstrings,''
Nucl.\ Phys.\ {\bf B276} (1986) 291.}
\lref\relevance{J.~Harvey, D.~Kutasov, E.~Martinec, ``On the Relevance of Tachyons,'' hep-
th/0003101.}
\lref\prei{C.~Preitschopf, C.~Thorn, S.~Yost, ``Superstring Field Theory,'' Nucl.\ Phys.\
{\bf B337} (1990) 363.}
\lref\yaone{I.~Ya.~Aref'eva, P.~Medvedev, A.~Zubarev, ``Background Formalism for 
Superstring
Field Theory,'' Phys.\ Lett.\ {\bf 240B} (1990) 356.}
\lref\yatwo{I.~Ya.~Aref'eva, P.~Medvedev, A.~Zubarev, ``New Representation for String Field
Solves the Consistency Problem for Open Superstring Field Theory,'' Nucl.\ Phys.\
{\bf B341} (1990) 464.}
\lref\pomarin{M.~Marino, ``On the BV Formulation of Boundary Superstring Field Theory,'' 
hep-th/0103089.}
\lref\gera{A.~Gerasimov, S.~Shatashvili, ``On Exact Tachyon Potential 
in Open String Field Theory,'' JHEP {\bf 0010} (2000) 034; hep-th/0009103.}
\lref\tseytfour{A.A.~Tseytlin, ``Sigma Model Approach to String Theory Effective Actions with
Tachyons,'' hep-th/0011033.}
\lref\andreev{O.~Andreev, ``Some Computations of Partition Functions and Tachyon Potentials
in Background Independent Off-Shell String Theory,'' hep-th/0010218.}

%%%%%%%%%%%%%%%%%%%%%%%%%%%%%%%%%%%%%%%%%%%%%%%%%%%%%%%%%

\noindent{\bf 1. INTRODUCTION} 
\medskip

Recently there has been a resurgence of interest in the so-called
boundary string field theory (BSFT) developed by Witten \refs{\witone,\wittwo}
and Shatashvili \refs{\shataone,\shatatwo}. The main reason is that
this version of open string field theory provides an intuitive
framework for the study of tachyon condensation in open string theory
\refs{\gera,\kutone,\kuttwo}. In contrast to the cubic string field theory, where
an infinite number of fields condense, BSFT allows for the tachyon field to 
condense alone, while the remaining open string tower continues to have 
vanishing
expectation values \relevance.

Tachyon condensation of unstable D-brane systems 
is a completely classical phenomenon described 
in BSFT by the RG flow of the tachyon perturbation
on the disk worldsheet. The
exact expressions for the tachyon potential 
and the descent relations among D-branes, as expected from \sen,
can be derived in a straightforward manner in this approach.

The idea behind BSFT is to consider the classical 
open string field theory as a theory
on the space of all boundary interactions on the disk with a fixed
conformal worldsheet action in the bulk. The implementation of this idea
relies on the use of the Batalin-Vilkovisky (BV) formalism \refs{\bataone,\batatwo}.
This construction was initially presented in \refs{\witone} for the bosonic
string and provides a generic expression for the spacetime action that can be
related to the disk partition function in a simple way
\foot{To derive this relation one must also assume that ghosts and matter are decoupled.}
\refs{\wittwo,\shataone,\shatatwo}
\eqn\bosact{
S_B = Z - \beta^i \frac{\partial}{\partial \lambda^i} Z
}
with $\beta^i$   
the worldsheet beta function for the boundary coupling $\lambda^i$.

In the absence of an analogous construction for the superstring, the authors of
\kuttwo \ conjectured that the corresponding spacetime action $S_F$ in the 
presence of worldsheet supersymmetry, should
be exactly equal to the disk worldsheet partition function $Z$. This conjecture
was motivated by a number of arguments involving the properties of $Z$ at the 
conformal points, its finiteness,
which is guaranteed by the presence of worldsheet supersymmetry, 
as well as similar proposals \refs{\tseytone,\tseyttwo,\tseytthree}
\foot{For a more recent discussion see also \tseytfour.}
in the context of a low-energy effective description. 
Further evidence for the validity of this choice was provided 
by the consistency of the results obtained from the analysis of
tachyon condensation in superstring theory \refs{\kuttwo,\larsen,\taka}.

In this note we complete the above picture by
repeating the analysis of \witone \ and 
applying the BV formalism on the space of supersymmetric
worldsheet boundary perturbations on the disk.   
Our formulation of boundary superstring field theory (SBSFT) is
cast entirely in superspace language. 
The crucial parts of the construction
involve the definition of 
a fermionic vector field and the definition of the appropriate BV antibracket.
The latter presents subtleties 
associated to the nature of the superconformal ghosts and the
existence of different ``pictures''. Nevertheless, an appropriate antibracket
exists and the BV formalism provides a corresponding spacetime action, 
which turns out to be exactly equal to the disk partition function, 
as was conjectured in \kuttwo.

The organization of this paper is as follows. In section 2 we provide
a brief review of the bosonic BSFT emphasizing the main points
of the construction. 
In section 3 we analyze the boundary superstring field theory in detail. We construct  
the appropriate BV structure and use it to
derive the equality $S_F = Z$. 
Moreover, since the spacetime
action is a monotonically decreasing function along the RG flow, 
we may identify it
with the boundary entropy on the disk.
We present a further check of these points by performing 
explicit calculations in conformal perturbation
theory up to third order. In section 4 we close with a short summary
of our results and a brief discussion on related issues.

As this paper was being prepared for submission a preprint appeared
\pomarin \ which addresses similar topics.  

%%%%%%%%%%%%%%%%%%%%%%%%%%%%%%%%%%%%%%%%%%%%%%%%%%%%%%%%%

\vskip .5cm
\noindent{\bf 2. REVIEW OF BOSONIC BSFT} 
\medskip

The worldsheet $\sigma$-model approach to string theory suggests that spacetime
fields should be viewed as generalized coupling ``constants'' of two-dimensional
worldsheet interactions. If we think of the string field as a collection of spacetime
fields, this picture implies that the string field simply encodes the data of a
two-dimensional field theory. Because of this, it is natural to expect that 
string field theory can be formulated on the ``space of all two-dimensional
field theories''. 

Boundary string field theory is an open string field theory based
on this idea. At the classical level its precise formulation relies on the 
application of the BV formalism
\foot{The application of the BV formalism has also been very useful in 
the formulation of the open cubic string field theory 
\refs{\bochi,\thorn} and closed string field theory \refs{\zwione}.} 
on the configuration space of all two-dimensional
field theories on the disk with arbitrary boundary interaction terms 
and a fixed conformal worldsheet action in the bulk.

In the following we review some basic features of the BV formalism 
and emphasize those that play a
prominent role in the construction of BSFT. For a more detailed exposition 
on the BV formalism we refer the reader to \refs{\gomis,\henneaux}.

One starts with a field configuration space $\cal M$ equipped with a supermanifold
structure, a non-degenerate closed odd two-form $\omega$ and a $U(1)$ symmetry under
which $\omega$ has charge -1. We call this charge (BV) ghost number. The 
two-form $\omega$ gives rise to an antibracket, in the same way that a symplectic form
gives rise to a Poisson bracket.
In local coordinates $u^I$, the antibracket of two functions $F$ and $G$ is given by
\eqn\anti{
\{ F , G \} = \frac{\partial_r F}{\partial u^I} \omega^{I J}
\frac{\partial_l G}{\partial u^J}. \ 
}

The focal point of the BV formalism is the $\it{master \ equation}$
\eqn\master{
\{ S , S \} = 0
}
for the master action $S$. This equation guarantees the gauge invariance of the system.
Notice that because of the fermionic
nature of $\omega$ the master equation is not trivially zero.

One way to satisfy the master equation identically is by choosing an appropriate 
fermionic vector field $V$ on $\cal{M}$ that has ghost number 1 and satisfies the equation
\eqn\defv{
i_V \omega = d S.
}
Borrowing the language of the Hamiltonian formalism, we would say that $V$ is the ``Hamiltonian
vector field'' of the action $S$.
As usual, $i_V \omega$ denotes the
inner product of the vector $V$ with the form $\omega$ and $d$ is simply the
exterior derivation on $\cal{M}$. 
The master equation will be satisfied identically if and only if $V$ is nilpotent,
i.e. $V^2=0$ and there exists at least one point where $V=0$ \witone. The latter is a very
natural requirement because the points of vanishing $V$ are precisely the extrema of
the spacetime action where the classical equations of motion are
satisfied.

Moreover, the exterior derivative of equation \defv\ gives
\eqn\dvomega{
d(i_V \omega)=0.
}
Given the closedness of the antibracket, it is not hard to show that this 
is presicely the statement that $\omega$ is invariant under infinitesimal transformations
generated by $V$, i.e. that
\eqn\lie{
{\cal L}_V \omega=(di_V +i_V d)\omega=0.
}
Hence, by the Poincare lemma, the property that the antibracket is $V$-invariant guarantees
the existence of an action $S$ given by \defv. This is true, of course, only locally on
the space of theories.

To summarize, the strategy for constructing a gauge invariant open string spacetime
action involves two steps. First we define a nilpotent fermionic vector
field $V$ of ghost number 1 that plays the role of the ``Hamiltonian'' vector field of 
the action $S$ and second we appropriately define a $V$-invariant odd symplectic
two-form $\omega$ with ghost number -1. 
The master action $S$ is then determined by \defv.  

This is the general setup. 
In the bosonic BSFT \witone \ the above construction works as follows.
The configuration space ${\cal M}_B$ consists, roughly speaking,
of all possible two-dimensional worldsheet theories with a fixed conformal action 
$\cal{S}_{\rm{bulk}}$
in the bulk of the disk (at the classical level) and a boundary action
$\cal{S}_{\rm{bdy}}$ with arbitrary
interactions on the boundary.
After choosing, for instance, the standard flat action for the bulk,
the dynamics of the two-dimensional field theories we want to consider are
given by 
\eqn\action{\eqalign{
{\cal S} &= {\cal S}_{\rm{bulk}} + {\cal S}_{\rm{bdy}} =
\cr  
\frac{1}{4 \pi \alpha'} \int_D d\sigma^1 d\sigma^2
\sqrt{h} \bigr( h^{a b} \partial_a X^{\mu} & \partial_b X_{\mu} \bigl) + 
\frac{1}{2 \pi} \int_D d\sigma^1 d\sigma^2 \sqrt{h} \; b^{ab} \partial_a c_b +
\frac{1}{2 \pi} \oint_{\partial D} d\tau {\cal V}
}}
where $D$ is the disk worldsheet, $h^{ab}$ a rotationally invariant metric 
and $\tau$ a periodic coordinate on the boundary $\partial D = S^1$.
Different points on the configuration space correspond to 
different choices of  
the boundary operator ${\cal V}$.

The vector field $V$ is associated to the
flow generated on ${\cal M}_B$ by the bulk BRST charge Q. Under this definition
it is straightforward to check that $V$ has the required properties. It is nilpotent and has
ghost number 1. Furthermore, when
the classical equations of motion are satisfied, BRST invariance is restored
and $V$ is vanishing as expected.

The construction of the odd symplectic two-form $\omega$ is more involved. \witone \ 
first defines $\omega$ on-shell and later extends the definition off-shell. This
extension, however, involves a subtle redefinition of the actual degrees of freedom of
the theory and leads to an enlargement of the actual space of theories. This subtle
point comes about in the following way.
Tangent vectors to an on-shell submanifold of ${\cal M}_B$
are spin one primary fields $\delta \cal{V}$. Since the BRST transformation
of such fields should leave the worldsheet action 
invariant we deduce that there
must be some operator $O$ of ghost number 1, such that
\eqn\onshe{
\{ Q,\delta \cal{V} \}=\sl \partial_{\rm \tau} O.
}
On-shell, $O$ can be uniquely determined \foot{Up to a total derivative
that has no effect on the worldsheet theory.}
from $\delta \cal V$. It satisfies the following
two equations 
\eqn\vo{
\delta {\cal V} =b_{-1} O
}
and
\eqn\brsclos{
\{ Q,O \}=0.
}

Off-shell, however, the second equation is no longer valid and the first one,
which still makes sense, 
determines $O$ only up to terms of the form
$b_{-1}(...)$. 
Since $O$ seems to be more fundamental than $\delta \cal V$ in some respects,
it was proposed in \witone \ to consider an enlarged space of theories determined not  
only by the worldsheet action \action \ but also by a ghost number 1
local operator $O$ satisfying equation
\vo. Thus, it seems more appropriate to view ${\cal M}_B$ as the space of 
the operators $O$ and
not as the space of the boundary perturbation operators ${\cal V}$.

Hence, given two
vectors $\delta_1 O$ and
$\delta_2 O$ at a point $O$ of the enlarged configuration space,
we define the odd symplectic form $\omega$ by
\eqn\anti{
\omega(\delta_1 O, \delta_2 O) = (-)^{\epsilon(\delta_1 O)}
\oint_{\partial D} d\tau_1 d\tau_2
\langle \delta_1 O(\tau_1)  \delta_2 O(\tau_2) \rangle}
with the correlation function being computed in the worldsheet theory
with boundary interaction ${\cal V}=b_{-1}O$.
This definition is slightly different from that of \witone\ by a sign factor. 
This factor is introduced in order to get the correct exchange property
\eqn\exchange{
\omega(\delta_1 O,\delta_2 O)=(-)^{(\epsilon_1+1)(\epsilon_2+1)+1}
\omega(\delta_2 O,\delta_1 O)
}
where $\epsilon_i=\epsilon(\delta_i O)$. For a similar definition in the context of
closed string field theory see for example \zwitwo.
Notice also that in the above expressions we still define the statistics 
of the arguments of $\omega$ as the natural statistics of the corresponding
$\delta {\cal V}$ fields.
For the precise off-shell definition of $b_{-1}$ and a proof
that \anti\ 
has the required properties, we refer the reader to \witone.

Now that we have established the needed BV structure we can easily write down
an expression for the master action $S_B$ (up to an irrelevant sign)
by using equation \defv\ and
the explicit form of the fermionic vector field $V(O) = \{Q, O\}$
\eqn\bsftone{
d S_B = \oint_{\partial D} d\tau_1 d\tau_2 \langle dO(\tau_1)  
\{ Q, O \}(\tau_2) \rangle.
}

Under the simplifying assumption that 
ghosts and matter are decoupled one can set 
$O=c{\cal V}$. In that case, using different approaches, the authors of
\wittwo \ and \refs{\shataone,\shatatwo} proved that 
\eqn\bosact{
S_B = Z - W^i \frac{\partial}{\partial \lambda^i} Z
}
where a generic expansion of the boundary operator  
${\cal V} = \sum_i \lambda^i {\cal V}_i$ has been implied.
$W^i$ is a vector field on ${\cal M}_B$. More precisely, 
it was identified up to second order in conformal perturbation theory
\refs{\shataone,\shatatwo} with the
beta function $\beta^i$, which  corresponds to the worldsheet 
RG flow of the coupling $\lambda^i$.

%\break

%%%%%%%%%%%%%%%%%%%%%%%%%%%%%%%%%%%%%%%%%%%%%%%%%%%%%%%
\vskip 0.5cm
\noindent{\bf 3. BOUNDARY SUPERSTRING FIELD THEORY} 

\

{\bf a. The BV formalism of SBSFT }
\medskip
 
Boundary superstring field theory is formulated on the space ${\cal M}_F$
of all
worldsheet supersymmetric two-dimensional field theories on the superdisk
with the usual NSR action in the bulk. In order to have a manifestly supersymmetric
formalism we use a superspace notation.
In particular, the worldsheet action takes the following form
\eqn\susyaction{\eqalign{
{\cal{S}}_{NSR} &= \cal{S}_{\rm{bulk}} + \cal{S}_{\rm{bdy}} =
\cr  
\frac{1}{4 \pi \alpha'} \int d^2 z \; d^2 \theta
D_{\bar{\theta}}{\bf X^{\mu}} D_{\theta}{\bf X_{\mu}}& + 
\frac{1}{2 \pi} \int d^2 z \; d^2 \theta B D_{\bar{\theta}} C + 
\int d\tau d\theta {\cal V}.
}}
Our conventions follow those of \refs{\polch,\fms} and for the ghost and
antighost superfields respectively  we set 
\eqn\C{
C(z,\theta) = c(z) + \theta \gamma(z)
}
\eqn\B{
B(z,\theta) = \beta(z) + \theta b(z).
} 
The bottom and upper components of the superfield boundary perturbation
will be generically given in the form
\eqn\boupert{ 
{\cal V}(\tau,\theta) = D(\tau) + \theta U(\tau).
}

For example, in the case of a single unstable non-BPS D9 brane in type IIA 
superstring theory, the
tachyon perturbation is given by the worldsheet action
\eqn\tachyon{
{\cal S}_{\rm bdy}=\int_{\partial D} d\tau d\theta (\Gamma D \Gamma+\Gamma T({\bf X}))
}
with $\Gamma=\eta+\theta F$ an auxiliary boundary fermion \refs{\kuttwo,\relevance}.
In that case  
\eqn\pertach{
{\cal V}=\Gamma D \Gamma+\Gamma T({\bf X}).
}  

Furthermore, we can express the superconformal ghosts 
in a bosonized form \fms 
\eqn\bosonization{
\beta(z) = e^{- \phi(z)} \partial\xi(z),  \;\;\;\;\;\;
\gamma(z) = e^{\phi(z)} \eta(z)
}
where $\xi$ is a fermion of dimension 0 and $\eta$ a fermion of dimension
1. Since we find this language very useful for the subsequent analysis, let us
briefly recall a few relevant facts. 

The zero-mode of $\xi$ does not enter the above bosonized expressions
and this results in a multiplicity of physically equivalent vacuum states that
lead to different irreducible representations of the superconformal
algebra, known as ``pictures''. 
A vertex operator
with a factor $e^{q \phi}$ is by definition in the $q$ picture.

When calculating on-shell amplitudes, pictures can be used in a more or less arbitrary manner
as long as the total superghost
number of the insertions is -2. In terms of the bosonized form of the ghosts
this condition implies a total $\phi$-charge -2.
For example, one may include two vertex operators in
the -1 picture and the rest in the 0 picture \fms. 
Switching between different pictures 
can be achieved by the use of the picture changing operator 
\eqn\pct{
X = Q \cdot \xi =
-\partial\xi c + e^{\phi} T_{F}^{m} - \partial\eta b e^{2 \phi} 
- \partial(\eta b e^{2\phi})
}
which {\it increases} the $\phi$-charge by 1 or the use of the inverse
picture changing operator \refs{\witipco, \belopo}
\eqn\ipct{
Y = - \partial\xi c e^{-2 \phi}
}
which {\it decreases} the $\phi$-charge by 1.

The freedom of moving a picture changing operator 
inside an amplitude is not, however, a valid off-shell
operation. This is an important subtlety of the superstring case and
must be taken into account in the following manipulations.

\

\

\break
%\vskip .5cm

{\it The definition of $V$}
\medskip

In complete analogy to the bosonic case, it is again natural to associate
the fermionic vector field $V$ to the flow generated
on ${\cal M}_F$ by the bulk BRST charge $Q$. The only extra subtlety in the superstring case
is that we choose ${\cal M}_F$ as a space of superfields. Hence, a sensible definition of
a vector field should generate flows that respect this property. More
precisely, this property is satisfied if and only if the generator of the corresponding flow
anticommutes with the generator of worldsheet SUSY. In our case, this is true
since
\eqn\qg{
\{Q,G_{-1/2}\}=0.
} 

Moreover, one can check that $V$ also inherits  
the rest of the required properties.
It is nilpotent, because $Q$ is nilpotent, and has ghost number 1.
We should emphasize that by ghost number we mean here the BV ghost number that coincides
with the total ghost and superghost number of the $bc$ and $\beta \gamma$ systems
respectively. The details of this definition of $V$ are the same as those of the bosonic case 
and we refer the reader to \witone.

%\break
\vskip .5cm

{\it The definition of $\omega$}

\medskip

We are looking for an appropriate two-form on the configuration space ${\cal M}_F$
of supersymmetric boundary perturbations on the superdisk. 
In order to grow the right intuition about this form we first
consider what happens on-shell. 
Let us also simplify
the situation further by assuming that ghosts and
matter are decoupled, so that the boundary interaction ${\cal V}=D+\theta U$ 
has no ghost dependence. In the $NS$ sector the -1 picture vertex operator
corresponding to $\cal V$ will be of the form 
\foot{The vertex operators given by $\Lambda$ correspond to the so-called 
strongly
physical states. For a related discussion see \belopo .}
\eqn\sps{
\Lambda =- c e^{- \phi} D.
}
We obtain a
0 picture representation of this operator by acting with the picture changing operator $X$
\eqn\spszero{
X \cdot \Lambda = \gamma D - cU.
}
This expression is precisely the upper component of the
superfield
\eqn\defg{
G = C {\cal V}
}
which is the natural supersymmetric generalization of the corresponding
bosonic expression $O=c{\cal V}$. We propose that $G$ should be considered as the fundamental
object of boundary superstring field theory. This is actually analogous
to the classical string field of the modified cubic superstring 
field theory \refs{\prei,\yaone,\yatwo} where the basic object is
a 0 picture, ghost number 1 operator. 

Given two superfield tangent vectors $\delta_1 {\cal V}$ and $\delta_2 {\cal V}$ 
we define the odd symplectic two-form $\omega$ in the superstring case
as a two-point function at the perturbed point ${\cal V}$ in the following way
\eqn\defomega{
\omega(\delta_1 G,\delta_2 G)=(-)^{\epsilon(\delta_1 G)}
\int d\tau_1 d\tau_2 d\theta_1 d\theta_2
\langle Y(\tau_1)\delta_1 G(\tau_1,\theta_1) Y(\tau_2) \delta_2 G(\tau_2,\theta_2)
\rangle }
with $\delta_1 G$ and $\delta_2 G$ given by \defg.
The main difference with respect to the bosonic definition are the
two insertions
of the inverse picture changing operator $Y$ in front of the 0 picture
superfields. Their presence is required because 
a non-vanishing 
expectation value should have a total superghost
number -2.

Expression \defg , however, is only valid
on-shell and only under the assumption of ghost-matter decoupling.
Nevertheless, we can rewrite it as
\eqn\bminusg{
{\cal V}=b_{-1} G.
}
As was explained in \witone, this form also makes sense off-shell. Hence, 
for the off-shell definition of the antibracket, we propose to consider the defining
relation \defomega \  but with tangent vectors $\delta_1 G$ and $\delta_2 G$ 
given now (implicitly) by \bminusg \ and not by \defg. 

Equation \bminusg \ does not define the superfield $G$ uniquely. 
As in the bosonic case, we circumvent this
problem by considering an enlarged space of theories ${\cal M}_F$
determined not only by the
worldsheet action \susyaction, but also by a 0 picture, ghost number 1 local superfield $G$
satisfying \bminusg. In some sense, we consider the superfields $G$ 
as the fundamental degrees of freedom of the theory. Nevertheless, we still define
the statistics of the arguments of $\omega$ as the natural statistics of the 
corresponding $\delta {\cal V}$ fields.

We also want to emphasize that
contrary to our on-shell experience from the first quantized 
description of string theory, the position of the picture changing operators 
in the definition of $\omega$ is important. 
We cannot move them freely inside the correlator because this is not a valid off-shell 
operation.

The next step is to verify that $\omega$ given by \defomega \ has all the required 
properties, i.e.
that it is a $V$-invariant non-degenerate odd symplectic two-form 
with BV ghost number -1.

%ghost number -1

Since the total ghost number of the vacuum is -1 (-3 from the 
$bc$ system and +2 from the $\beta \gamma$ system) and the ghost number of both  
insertions $Y \cdot \delta_i G$ is 0, we deduce that
$\omega$ has BV ghost number -1, as expected.  

%odd 

The statistics $\epsilon(\omega)$, on the other hand, is given by the sum 
\eqn\statis{
\epsilon(\delta_1 G)+
\epsilon(\delta_2 G) + 2 \epsilon(Y)+2=\epsilon(\delta_1 G)+\epsilon(\delta_2 G) \ \ 
\rm{(mod \ 2)}
}
with 
the extra +2 in the left-hand side coming from the two $\theta$ integrations.
Since a non-vanishing correlator requires $\epsilon(\delta_1 G)+\epsilon(\delta_2 G)
=1 \ \ \rm{(mod \ 2)}$ we get
\eqn\odd{
\epsilon(\omega)=1 \ \ \rm (mod \ 2)
}
and therefore $\omega$ is odd.

%nondegenerate

In order to show the non-degeneracy of the antibracket let us go on-shell. 
In that case $\omega$ 
vanishes for BRST exact insertions and therefore we may regard it as a
two-form on the space of classical solutions. It is non-degenerate because it is
related to the Zamolodchikov metric $g$ on the space of conformal field theories. We can
prove this by setting $\delta_1 G=C{\cal V}$ and $\delta_2 G=C\partial C W$, with
$V$ and $W$ two spin 1/2 primary matter superfields. It follows that
$\omega(\delta_1 G,\delta_2 G) \propto g(D_V,D_W)$, where
$D_V$ and $D_W$ are the bottom components of the matter superfields. 
Thus, the non-degeneracy of $\omega$ follows from the non-degeneracy of the Zamolodchikov
metric in complete analogy to the bosonic situation \witone.

%%%%%%%%%%%%%%%%%%%%%%%%%%%%%%%%%%%%%%%%%%%%%%%%%%%%%%%%%%%%%%%%%%%%%%%%%%%%%%

To prove that $d \omega = 0$, let us introduce 
local coordinates
$\lambda^i$ on ${\cal M}_F$. We can then write a generic superfield tangent vector 
$\delta G$ in terms of the expansion 
$\delta G = \sum_i \lambda^i \delta_i G$. By definition, we have
\eqn\domega{
d\omega(\delta_i G, \delta_j G, \delta_k G) = 
\frac{\partial}{\partial \lambda^i} \omega(\delta_j G, \delta_k G) \pm {\rm (cyclic \ 
permutations).}
}
The derivative with respect to $\lambda^i$ gives
\eqn\deriv{
\frac{\partial}{\partial \lambda^i} \omega (\delta_j G, \delta_k G) = 
\int \prod_{\beta=1}^3 d\tau_{\beta} d\theta_{\beta} 
(-)^{\epsilon_j}
\bigg \langle
\bigg ( b_{-1} \delta_i G(\tau_1,\theta_1)\bigg ) Y(\tau_2) \delta_j G(\tau_2,\theta_2) 
Y(\tau_3) \delta_k G(\tau_3,\theta_3) \bigg \rangle 
}
with $\epsilon_i = \epsilon(\delta_i G)$.
Hence, written explicitly, equation \domega \ takes the form
\eqn\domega{\eqalign{
& d\omega(\delta_i G, \delta_j G, \delta_k G) = (-)^{\epsilon_j}
\int \prod_{\beta=1}^3 d\tau_{\beta} d\theta_{\beta} \bigg (
\bigg \langle (b_{-1}\delta_i G) (Y \delta_j G) (Y \delta_k G) \bigg \rangle \cr
+ (-& 1)^{\epsilon_i} 
\bigg \langle (Y\delta_i G) (b_{-1} \delta_j G) (Y \delta_k G) \bigg \rangle
+(-1)^{\epsilon_i+\epsilon_j} 
\bigg \langle (Y\delta_i G) (Y \delta_j G) (b_{-1} \delta_k G) \bigg \rangle
\bigg ).
}}

The fact that the above expression vanishes follows from 
the $Q$ invariance of the unperturbed correlators and the 
invariance under $b_{-1}$.
The details of the relevant calculation can be found in appendix A.

The final property we have to check is $V$-invariance, i.e. 
$d(i_V \omega) = 0$. More explicitly, one must show that
\eqn\invone{
\frac{\partial}{\partial \lambda^i} 
\int \prod_{\beta=1}^3 d\tau_{\beta} d\theta_{\beta} (-)^{\epsilon_j}
\bigg \langle Y(\tau_1) \delta_j G(\tau_1,\theta_1) 
Y(\tau_2) [Q , G ] (\tau_2,\theta_2) \bigg \rangle
\pm (i \leftrightarrow j)=0. 
}
For comments on this proof we refer the reader again
to appendix A.

%%%%%%%%%%%%%%%%%%%%%%%%%%%%%%%%%%%%%%%%%%%%%%%%%%%%%%%%%%%%%%%%%%%%%%%%%%%%%%%

This concludes our discussion of the BV structure of the boundary superstring field
theory. We now employ the above formalism to investigate the master action.

%\break

\vskip 0.5cm 
{\bf b. The relation between $S_F$ and $Z$}

\medskip

The spacetime action of SBSFT (again up to an irrelevant sign factor) follows directly 
from equation \defv, the definition of the vector field $V$ and the definition of the 
odd symplectic form $\omega$ 
\eqn\faction{
dS_F=\int d\tau_1 d\tau_2 d\theta_1 d\theta_2 
\langle Y(\tau_1)dG(\tau_1,\theta_1)Y(\tau_2)[ Q,G ] (\tau_2,\theta_2) \rangle.
}
Furthermore, since we make the assumption that ghosts and matter are decoupled, we may set
$G=C\cal V$, in which case the action takes the form
\eqn\saction{
dS_F=\int d\tau_1 d\tau_2 d\theta_1 d\theta_2
\langle Y(\tau_1) C(\tau_1,\theta_1)d{\cal V}(\tau_1,\theta_1) Y(\tau_2)[ Q,C
{\cal V} ](\tau_2,\theta_2)
\rangle.
}
In the following discussion we perform an explicit calculation 
of this expression and
demonstrate how it relates to the superdisk partition function $Z$.

Let us start by making the simplifying assumption that ${\cal V}$ has a definite 
scaling dimension $h$. The commutator $[ Q,C{\cal V}]$ is given more explicitly by
\eqn\com{
[ Q,C{\cal V} ]=\{Q,C\}{\cal V}-C\{Q,{\cal V}\}.
}
Moreover, the following two equations hold \fms
\eqn\qc{
\{ Q,C \}=C\partial_{\tau} C-\frac{1}{4}(D_{\theta}C)(D_{\theta}C)
}
\eqn\qv{
\{Q,{\cal V}\}=C\partial_{\tau}{\cal V}-\frac{1}{2}(D_{\theta}C)(D_{\theta}
{\cal V})+h(\partial_{\tau}C){\cal V}.
}
Substituting them back into equation \com \ gives
\eqn\qcv{
[Q, C{\cal V}]=[(1-h)C\partial_{\tau}C-\frac{1}{4}(D_{\theta}C)(D_{\theta}C)]{\cal V}+
\frac{1}{2}C(D_{\theta}C)(D_{\theta}{\cal V}).
} 
An explicit calculation of the normal ordered expression $Y[Q,C{\cal V}]$ 
in components (see appendix B for details) gives 
\eqn\yqcv{
Y[Q,C{\cal V}]=
\bigg ( (1-h) Yc\partial_{\tau}c-\frac{1}{4}Y\gamma^2 \bigg ) {\cal V}+
\frac{1}{2}\theta (Y\gamma^2-Yc\partial_{\tau}c)D_{\theta} {\cal V}+
\bigg( h-\frac{1}{2}\bigg) \theta c\partial_{\tau}ce^{-\phi}D.
}
Non-vanishing amplitudes require three insertions of the $c$ ghost
and therefore only the last term in the above expression contributes to 
the action. Hence,
\eqn\caction{
dS_F=\bigg( \frac{1}{2}-h \bigg ) \oint_{\partial D}  d\tau_1 d\tau_2 
\langle c(\tau_1)c(\tau_2)\partial_{\tau} c(\tau_2) \rangle_{bc}
\langle e^{-\phi(\tau_1)}e^{-\phi(\tau_2)} \rangle_{\beta \gamma}
\langle dD(\tau_1)D(\tau_2) \rangle_m .
}
Since
\eqn\ccor{
\langle c(\tau_1)c(\tau_2)\partial_{\tau_2}c(\tau_2) \rangle=
2(\cos(\tau_1-\tau_2)-1)=-4\sin^2 \bigg ( \frac{\tau_1-\tau_2}{2} \bigg)
}
and
\eqn\betacor{
\langle e^{-\phi(\tau_1)}e^{-\phi(\tau_2)} \rangle = \frac{1}{2} \frac{1}
{\sin(\frac{\tau_1-\tau_2}{2})}
}
we take
\eqn\taction{
dS_F=(2h-1) \oint_{\partial D} d\tau_1 d\tau_2 \sin \bigg( \frac{\tau_1-\tau_2}{2} \bigg )
\langle dD(\tau_1)D(\tau_2) \rangle.
}

For a generic perturbation $\cal V$ parametrized by couplings $\lambda^i$
and operators ${\cal V}_i$ of conformal weight $h_i$
\eqn\expan{
{\cal V}=\sum_i \lambda^i {\cal V}_i
}
and the above equation becomes
\eqn\anaction{
\frac{\partial S_F}{\partial \lambda^i}=\bigg (h_j - \frac{1}{2} \bigg)
\lambda^j G_{ij}(\lambda)
}
where
\eqn\metr{
G_{ij}=2 \oint_{\partial D} d\tau_1 d\tau_2 \sin \bigg (\frac{\tau_1-\tau_2}{2} \bigg ) 
\langle D_i(\tau_1)D_j(\tau_2) \rangle.
}
These equations should be compared to equations (2.9) and (2.10) of reference 
\kutone \ 
for the bosonic case.

%%%%%%%%%%%%%%%%%%%%%%%%%%%%%%%%%%%%%%%%%

For a unitary theory, the integral appearing in 
equation \metr\ is expected to give a positive definite metric
$G_{ij}$. Indeed, the Hilbert space of a unitary theory should have, by
definition, a positive definite norm given by the time ordered expectation
value. In particular, for a Hilbert space of
odd excitations, this statement implies that we should be able to
write an expectation value 
of the form $\langle D(x_1) D(x_2) \rangle$ as ${\rm sign}(x_1-x_2) f_D
(x_1-x_2)$, with $f_D$ being a positive function. Combining this extra $\rm
sign$ factor with the sine in the integral of \metr\ gives a manifestly
positive form to the metric $G_{ij}$.

Also, equation \anaction \ cannot be correct in
general because
$(\frac{1}{2}-h_j)\lambda^j$ is not a covariant expression on the space of theories.
The relevant argument goes exactly the same way as in the bosonic case \kutone.
The correct covariant generalization of \anaction\ is given by
\eqn\oneaction{
\frac{\partial S_F}{\partial \lambda^i}=\beta^j G_{ij}.
}
Thus, along the RG flow 
\eqn\rg{
\frac{\partial S_F}{\partial \log t}=-\beta^i \frac{\partial S_F}{\partial \lambda_i}=
-\beta^i \beta^j G_{ij}
}
where $t$ is an RG (length) scale parameter. 
Since $G_{ij}$ is positive definite, $S_F$ is a monotonically decreasing function
and also stationary at the conformal points. Furthermore, as we show
in the following discussion, $S_F$ equals to the disk partition
function $Z$. Therefore, $S_F$ can be identified with the boundary entropy and the above
analysis agrees very nicely with the conjecture of \refs{\affone,\afftwo}
in the context of boundary CFT.

In order to prove the conjectured relation $S_F=Z$, we make use of the 
``two-systems'' approach 
introduced in \wittwo. According to this, we assume that the matter system
consists of two decoupled subsystems with partition functions $Z_1$ and $Z_2$. 
Thus, the combined matter partition function $Z$ equals 
the product $Z_1 Z_2$ and the 
expansion \expan \ takes the form
\eqn\expantwo{
{\cal V}=\sum_i \lambda^i V_i+\sum_{k} \rho^{k} \tilde{V}_{k}
}
with $\lambda^i$ couplings for the first system and $\rho^{k}$ couplings for the
second system. Accordingly, \taction \ becomes
\eqn\expaction{
dS_F=\oint_{\partial D} d\tau_1 d\tau_2 \sin \bigg ( \frac{\tau_1-\tau_2}{2} \bigg ) 
\bigg \langle dD(\tau_1) \bigg ( \sum_i (2h_i-1) \lambda^i D_i (\tau_2)+
\sum_{k} (2h_k-1) \rho^{k} \tilde{D}_{k}(\tau_2)\bigg ) \bigg \rangle.
}

After substituting into this equation the full expansion of $dD$ we get two kinds of terms.
One involves two point functions of the form $\langle D_iD_j \rangle$ and 
$\langle \tilde{D}_{k}\tilde{D}_{l} \rangle$. The other involves mixed terms of the
form $\langle D_i \tilde{D}_{k} \rangle$. 
Since the two systems are decoupled these mixed
correlators factorize into a product of two one-point functions, which are identically
zero because the bottom components $D$ are fermionic. Moreover, even if these terms
were not zero, the resulting expression would involve the
integral of $\sin (\frac{\tau_1-\tau_2}{2} )$ over the circle and hence it would still be
vanishing.
The bosonic case, on the other hand, involved several non-vanishing
mixed terms and these were responsible
for the extra term with the $\beta$ function on the right side of \bosact . 
Thus, we find that 
\eqn\sexpnaction{\eqalign{
dS_F= \oint_{\partial D} d\tau_1 d\tau_2 \sin & \bigg (\frac{\tau_1-\tau_2}{2} \bigg )
\bigg ( \sum_{i,j} (2h_i-1) \lambda^i d\lambda^j \langle D_j(\tau_1)D_i(\tau_2) \rangle+
\cr
+& \sum_{k,l} (2h_l-1) \rho^{l}d\rho^{k} \langle \tilde{D}_{k}(\tau_1)\tilde{D}_{l}
(\tau_2) \rangle \bigg ).
}} 

After setting
\eqn\aform{
a=\sum_j a_j(\lambda)d\lambda ^j
=\sum_j d\lambda^j \bigg [ \oint_{\partial D} d\tau_1 d\tau_2 \sin \bigg 
( \frac{\tau_1-\tau_2}{2} \bigg )
\sum_i (2h_i-1)\lambda^i \langle D_j(\tau_1) D_i(\tau_2) \rangle_1 \bigg ]
}
\eqn\atilform{
\tilde{a}=\sum_{k} \tilde{a}_{k}(\rho)d\rho^{k}
=\sum_{k} d\rho^{k} \bigg [ \oint_{\partial D} d\tau_1 d\tau_2 \sin \bigg 
( \frac{\tau_1-\tau_2}{2} \bigg )
\sum_{l} (2h_l-1)\rho^{l} \langle \tilde{D}_{k}(\tau_1)\tilde{D}_{l}(\tau_2) \rangle_2 \bigg]
}
we can write equation \sexpnaction \ in a compact form as
\eqn\comaction{
dS_F=a Z_2+Z_1 \tilde{a}.
}

The one-forms $a$ and $\tilde{a}$ can be related to the partition functions $Z_1$ and
$Z_2$ by using the fact that $d^2 S_F=0$. Setting to zero the coefficient 
of $d\lambda^i \wedge d\rho^{k}$ in the expression of $d^2S_F$ gives
\eqn\arela{
dZ_1 \tilde{a} - a dZ_2=0.
}
Since the two systems are decoupled, this 
implies the existence of a non-zero constant $g$ such that
\eqn\arez{
a=gdZ_1, \ \ \ \tilde{a}=gdZ_2.
}
$g$ cannot be zero because $\omega$ is non-degenerate. In fact, as we show in
the next section, conformal perturbation theory fixes the value of $g$ to 1.
Thus, putting everything together gives 
\eqn\szo{
dS_F=d(Z_1Z_2)
}
or
\eqn\sz{
S_F=Z.  
}

The above derivation of the equality between the master action and the 
superdisk partition function was carried under the assumption that the matter
system consists of two decoupled subsystems. However, as pointed out in \wittwo ,
this restriction is not necessary. One can consider any matter system and carry
out the above analysis by adding an auxiliary decoupled system. At the end of
the calculation the auxiliary system can be suppressed by setting its couplings
to a fixed value.

\vskip .5cm

%%%%%%%%%%%%%%%%%%%%%%%%%%%%%%%%%%%%%%%%%%%%%%%%%%%%%%%
\noindent{\bf c. Some calculations in conformal perturbation theory}
\medskip

In the previous section we used a general argument to show the relation
\eqn\szg{
\partial_i S_F=g \partial_i Z
}
where $g$ is a constant factor. 
We examine now this relation from the point of view of conformal
perturbation theory up to 3rd order on the bare couplings $\lambda^i$ 
of the boundary perturbations. 
In particular, we consider the following expansion of equations \anaction \
and  \metr\ 
around the conformal point 
\eqn\sexpan{\eqalign{
\partial_i S_F &=2(h_j-\frac{1}{2}) \bigg ( \lambda^j \oint
d\tau_1 d\tau_2  
\sin \frac{\tau_1-\tau_2}{2} \langle D_i (\tau_1) D_j (\tau_2) \rangle _0 +
\cr &
\lambda^j \lambda^k \oint d\tau_1 d\tau_2 d\tau_3 \sin \frac{\tau_1-\tau_2}{2}
\langle D_i(\tau_1) D_j(\tau_2) U_k (\tau_3) \rangle_0 \bigg )
}}
and compare it to the corresponding expansion of the disk partition function  
\eqn\zexpan{
\partial_i Z=\lambda^j \oint d\tau_1 d\tau_2 \langle U_i(\tau_1) U_j(\tau_2) \rangle_0
+ \frac{1}{2} \lambda^j \lambda^k \oint d\tau_1 d\tau_2 d\tau_3 
\langle U_i(\tau_1) U_j(\tau_2) U_k(\tau_3) \rangle _0.
}
According to \szg, we should be able to verify the following 
two equations
\eqn\expanone{
2(h_j-\frac{1}{2}) \oint d\tau_1 d\tau_2 \sin \frac{\tau_1-\tau_2}{2}
\langle D_i(\tau_1) D_j(\tau_2) \rangle_0 =
g \oint d\tau_1 d\tau_2 \langle U_i (\tau_1) U_j (\tau_2) \rangle _0
}
in second order, and
\eqn\expantwo{\eqalign{
2(h_j-\frac{1}{2}) \oint d\tau_1 d\tau_2 d&\tau_3 \sin \frac{\tau_1-\tau_2}{2} 
\langle D_i(\tau_1) D_j(\tau_2) U_k(\tau_3) \rangle_0 + (j \leftrightarrow k) =
\cr &
g \oint d\tau_1 d\tau_2 d\tau_3 \langle U_i(\tau_1) U_j(\tau_2) U_k(\tau_3) \rangle_0.
}}
in third order.

For the second order computation, the explicit form of the correlators is as follows
\eqn\ddcor{
\langle D_i (\tau_1) D_j(\tau_2) \rangle_0 = \frac{c \delta_{h_i,h_j}}{
\bigg | \sin \frac{\tau_1-\tau_2}{2} \bigg |^{2h}} {\rm sign}(\tau_1-\tau_2),
}
\eqn\uucor{
\langle U_i (\tau_1) U_j (\tau_2) \rangle_0 = \frac{b \delta_{h_i,h_j}}{
\bigg | \sin \frac{\tau_1-\tau_2}{2} \bigg |^{2h+1}},
}
with $h=h_i=h_j$. After evaluating the relevant integrals by using the general 
expression
\eqn\sinint{
\int_0^{2\pi} \frac{d\tau_1}{2 \pi} \frac{d\tau_2}{2\pi} \bigg | \sin \frac{\tau_1-\tau_2}{2} \bigg |^z =
\frac{1}{\sqrt{\pi}}  \frac{\Gamma (\frac{1}{2}+\frac{z}{2})}{\Gamma (1+\frac{z}{2})},
} 
equation \expanone\ becomes
\eqn\cb{\eqalign{
2 (h-\frac{1}{2}) c \frac{\Gamma (1-h)}{\Gamma (\frac{3}{2} - h)}& = 
gb \frac{\Gamma (-h)}{\Gamma (\frac{1}{2}
-h)} \Rightarrow 
\cr 
2 c h &= g b.
}}  
This equation, with $g=1$, is a consequence of the SUSY Ward identities on the worldsheet. 
For a short derivation see appendix C.

Similarly, in third order we have the correlators 
\eqn\dducor{
\langle D_i (\tau_1) D_j (\tau_2) U_k(\tau_3) \rangle_0 = 
\frac{C_{ijk} \ \ {\rm sign}(\tau_1-\tau_2)}{\bigg | \sin \frac{\tau_1-\tau_2}{2} \bigg |^{
h_i+h_j-h_k-\frac{1}{2}}
\bigg | \sin \frac{\tau_2-\tau_3}{2} \bigg |^{h_j+h_k-h_i+\frac{1}{2}}
\bigg | \sin \frac{\tau_1-\tau_3}{2} \bigg |^{h_i+h_k-h_j+\frac{1}{2}} },
}
\eqn\uuucor{
\langle U_i(\tau_1) U_j(\tau_2) U_k (\tau_3) \rangle_0 =
\frac{B_{ijk}} 
{\bigg | \sin \frac{\tau_1-\tau_2}{2} \bigg |^{h_i+h_j-h_k+\frac{1}{2}}
\bigg | \sin \frac{\tau_2-\tau_3}{2} \bigg |^{h_k+h_j-h_i+\frac{1}{2}}
\bigg | \sin \frac{\tau_1-\tau_3}{2} \bigg |^{h_i+h_k-h_j+\frac{1}{2}} }.
}
By using the general expression
\eqn\threesinint{\eqalign{&
\int_0^{2\pi}  \frac{d\tau_1}{2 \pi} \frac{d\tau_2}{2\pi}  \frac{d\tau_3}{2\pi}
\bigg | \sin \frac{\tau_1-\tau_2}{2} \bigg |^a
\bigg | \sin \frac{\tau_2-\tau_3}{2} \bigg |^b
\bigg | \sin \frac{\tau_1-\tau_3}{2} \bigg |^c 
\cr 
= &\frac{1}{\pi ^{3/2}} 
\frac{ \Gamma (\frac{1}{2}(1+a)) \Gamma (\frac {1}{2} (1+b)) \Gamma (\frac{1}{2} (1+c))
\Gamma (1+\frac{1}{2}(a+b+c))}
{\Gamma (1+\frac{1}{2}(a+b)) \Gamma(1+\frac{1}{2}(b+c)) \Gamma(1+\frac{1}{2}(a+c))}.
}}
we can evaluate all terms in \expantwo . The result of this calculation for $g=1$ is 
\eqn\cbijk{
-2B_{ijk}\bigg(\frac{1}{2}-h_i\bigg)=\bigg(\frac{1}{2}-h_i-h_j-h_k\bigg) 
\bigg(C_{ijk} \bigg(\frac{1}{2}+h_k-h_i-h_j\bigg
)+C_{ikj}\bigg (\frac{1}{2}+h_j-h_i-h_k\bigg ) \bigg).
}
Since $C_{ijk}=C_{ikj}$, the above equation becomes
\eqn\cbijksym{
B_{ijk}=-\bigg(\frac{1}{2}-h_i-h_j-h_k\bigg) C_{ijk}.
}  
Again, we can verify this relation, as well as the symmetry properties of the
constants $C_{ijk}$, by using the SUSY Ward identities. The
relevant details can be found in appendix C.

Another important statement of the previous section was the identification of the worldsheet
partition function $Z$ with the boundary entropy of \refs{\affone,\afftwo}. 
We can check this identification by an
explicit conformal perturbation theory calculation,
 showing the decrease of the partition function 
under the renormalization group flow. We perturb the worldsheet action by 
some relevant operator very close to marginality and as a result the worldsheet theory  
flows towards a nearby conformal fixed point, where we
can still use perturbation theory to calculate the new value of the 
partition
function. A similar calculation can be found in appendix E 
of reference \afftwo.  

Thus, consider the boundary perturbation $\cal V$ with conformal weight $h=1/2-y$
and
$0<y \ll 1/2$. For simplicity, we assume that the RG flow is closed 
under this perturbation and that there is no mixing with other fields. 
The corresponding beta function up to second order equals
\eqn\betafun{
\beta(\lambda) =- y \lambda -\frac{B}{\pi} \lambda^2, 
}
with $B=B_{\cal VVV}$ the 3-point function constant appearing in \uuucor. 
This $\beta$ function implies the existence of a nearby fixed point given by
\eqn\fixedpoi{
\beta(\lambda^*)=0 \Rightarrow \lambda^*=-\frac{ y \pi}{ B} \ll 1.
}

In order to verify \betafun , consider an ultraviolet (length scale) 
cut-off $l$, an RG length scale parameter $t$ and the
boundary perturbation for bare coupling $\lambda$ written in the form
\eqn\bdyact{
S_{\rm bdy}=\bigg (\frac{t}{l}\bigg )^{y} \lambda \oint \frac{d\tau}{2\pi} U(\tau).
}
In this relation, $\lambda$, $\tau$ and $U$ are dimensionless. 

Expanding the partition function to quadratic order in the coupling and using
the OPE
\eqn\ope{
U (\tau_1) U(\tau_2) \sim \frac{B}{|\sin \frac{\tau_1-\tau_2}{2} |^{h+1/2}} 
  U(\tau_2)  }
gives
\eqn\betacalcone{\eqalign{&
e^{S_{\rm bdy}}=1+ \bigg (\frac{t}{l}\bigg)^{y} \lambda \oint \frac{d\tau}{2\pi} U(\tau) +
\frac{1}{2}  \bigg (\frac{t}{l}\bigg)^{2y} \lambda^2 \oint \frac{d\tau_1}{2\pi} \frac{d\tau_2}{2\pi} 
U (\tau_1) U(\tau_2) 
\cr &
=1+\bigg (\frac{t}{l}\bigg)^{y} \lambda \oint \frac{d\tau}{2\pi} U(\tau) +
\frac{1}{2} \bigg(\frac{t}{l}\bigg)^{2y} \lambda^2 \oint \frac{d\tau_1}{2\pi} U (\tau_1)
\oint \frac{d\tau_2}{2\pi} \frac{B}{|\sin \frac{\tau_1-\tau_2}{2} |^{h+1/2}} 
\cr &
=1+\bigg(\frac{t}{l}\bigg)^{y} \lambda \oint \frac{d\tau}{2\pi} U(\tau) +
\frac{1}{2} \bigg(\frac{t}{l}\bigg)^{2y} \lambda^2 \frac{B}{\sqrt{\pi}}
\frac{\Gamma (\frac{y}{2})}{\Gamma (\frac{1+y}{2})} \oint  \frac{d\tau}{2\pi}U(\tau)
\cr &
\sim 1+\bigg(\frac{t}{l}\bigg)^{y} \lambda \oint \frac{d\tau}{2\pi} U(\tau) +
\bigg(\frac{t}{l}\bigg)^{2y}  \lambda^2 \frac{B}{y \pi}
\oint  \frac{d\tau}{2\pi}U(\tau).
}}
In the last step we have set $y \sim 0$. Considering now these terms 
as a correction to
the initial perturbation, we find
\eqn\betacalctwo{
\frac{\delta \lambda(t)}{\delta \ln t}=y \lambda(t) + \frac{B}{\pi} \lambda(t)^2,
}
which gives precisely the beta function \betafun.

In terms of the bare coupling $\lambda=\lambda(l)$,
the partition function up to 3rd order is given by 
\eqn\zthird{
Z=1+\frac{1}{2} \bigg(\frac{t}{l}\bigg)^{2y} \lambda^2 \oint \frac{d\tau_1}{2\pi} \frac{d\tau_2}{2\pi} 
\langle U(\tau_1) U(\tau_2) \rangle_0 + 
\frac{1}{6} \bigg(\frac{t}{l}\bigg)^{3y} \lambda^3 \oint \frac{d\tau_1}{2\pi} 
\frac{d\tau_2}{2\pi} \frac{d\tau_3}{2\pi}
\langle U(\tau_1) U(\tau_2) U(\tau_3) \rangle_0.
}
After substituting the CFT expressions \uucor\ (with normalization $b=1$) 
and \uuucor\ 
we obtain
\eqn\zthirdone{
Z=1+\frac{1}{2} \bigg (\frac{t}{l}\bigg )^{2y} \lambda^2 
\frac{1}{\sqrt{\pi}}
\frac{\Gamma(y-\frac{1}{2})}{\Gamma(y)}+
\frac{1}{6}  \bigg (\frac{t}{l}\bigg )^{3y} \lambda^3 B  
\frac{1}{\pi \sqrt{\pi}} 
\frac{(\Gamma (\frac{1}{2}y))^3 \Gamma(-\frac{1}{2}+\frac{3}{2}y)}{(\Gamma(y))^3},
}
which in the $y \sim 0$ limit simplifies to
\eqn\zthirdtwo{
Z=1-y\lambda^2  \bigg (\frac{t}{l}\bigg )^{2y} -\frac{8}{3\pi}  B \lambda^3 
\bigg (\frac{t}{l} \bigg )^{3y}.
}

This equation can be re-expressed in terms of the renormalized coupling 
$\lambda (t)$ by solving the $\beta$ function equation \betacalctwo . 
The solution gives
\eqn\solbeta{
\lambda=\bigg ( \frac{t}{l} \bigg)^{-y} \frac{\lambda(t)}{1-\frac{\lambda(t)}{\lambda^*}
(1-(\frac{t}{l})^{-y})}.
}
Expanding this expression up to second order in $\lambda(t)$ and 
substituting the result into \zthirdtwo\ gives 
\eqn\zthirdthree{
Z=1- y \lambda(t)^2-\frac{2}{3\pi} \lambda(t)^3 B.
} 
Thus, in the IR limit where $\lambda(t) \rightarrow \lambda^*$, 
the total change
of the disk partition function between the UV and IR fixed points becomes
\eqn\deltaz{
\delta Z=-\frac{\pi^2 y^3}{3B^2}.
}
This result was also obtained in \afftwo.
As expected, we find that $Z$ decreases under the
renormalization group flow.

\vskip .5cm
%%%%%%%%%%%%%%%%%%%%%%%%%%%%%%%%%%%%%%%%%%%%%%%%%%%%%%%

\noindent{\bf 5. DISCUSSION}
\medskip

The aim of the above discussion was to show that on a formal level the construction of 
the (classical) boundary string field theory for the $NS$ sector of the superstring is more
or less parallel to the analogous construction of the bosonic case. However,  
in the superstring case there are certain extra subtle points. 
The first is the requirement to preserve worldsheet supersymmetry. 
We satisfy it by using a superspace formalism that is manifestly
supersymmetric. 
%we should stress that in our case ($NS$ boundary interaction terms on the disk)
%the global worldsheet supersymmetry is ``softly'' broken as a consequence of the
%antiperiodic boundary conditions on the disk \tseytone. 
A second subtlety in the superstring
case arises  from the involved nature of the superconformal ghosts. Vertex operators can be
chosen in different pictures and this choice involves the insertion of appropriate 
picture changing operators inside the correlation functions.

After defining the appropriate BV structure the spacetime action is determined by
\faction. Under the additional assumption of ghost and matter decoupling, this action takes 
a simpler and more appropriate for calculations form. Using the two-systems analysis of
\wittwo \ and conformal perturbation theory up to third order
we verified that the master action is non other than the disk partition function, 
exactly as it was conjectured in \kuttwo. 
This identification also suggests that the spacetime action can also be thought of as the
boundary entropy of \refs{\affone,\afftwo}. It takes the right value at the conformal points
and decreases along the RG flow.

Nevertheless, this construction of boundary superstring field theory is not at all complete.
First of all, the above analysis has to be extended appropriately to include
the $R$ sector. Secondly, the presented formalism
is plagued by the same problems that characterize the analogous construction of the
bosonic case. Very simply put, the construction is too formal. The space of all theories 
(with varying local boundary interaction terms) gives rise to serious ultraviolet divergences,
especially when one tries to add non-renormalizable boundary interaction terms.
These terms correspond to higher massive excitations of the open string and they
certainly
have to be included in any acceptable formulation of string field theory.

In order to
tackle these divergences an appropriate cut-off has to be introduced. The cut-off 
should respect the rotational invariance, the $b_{-1}$ Ward identities and 
$V$-invariance (i.e. the invariance under the closed BRST charge $Q$)
\foot{Given the previous invariances, $V$-invariance is equivalent to the 
statement that the boundary interaction does not modify the BRST charge. See
\refs{\shatatwo,\liwit} for further comments on this point.}.
In the superstring case it should also respect worldsheet supersymmetry. A cut-off
can be chosen to
respect all of the above symmetries except for $V$-invariance. Because of
this, at the end of the calculation one would like to remove the cut-off in
such a way that the $V$-invariance of the antibracket is restored. The relevant discussion
of \liwit \ for certain integrable boundary interactions in the bosonic case revealed that  
the removal of the cut-off presented difficulties. It is not clear however whether this
poses an insurmountable obstacle in making sense of the notion of a space of open string
theories with local boundary interactions. One expects similar 
difficulties in the superstring case as well.

Because of such problems the
question of whether this formalism can provide a rigorous and full formulation
of open string field theory is still open.
It might be possible, however, that a more careful application of the BV formalism could
provide the needed resolution of the above subtleties.

%%%%%%%%%%%%%%%%%%%%%%%%%%%%%%%%%%%%%%%%%%%%%%%%%%%%%%%%%%%%%%%%%%%%%%%%%

\vskip 1cm
\noindent{\bf Acknowledgments}

\

We would like to thank I. Ellwood, A. Hanany, D. Kutasov, F. Larsen, A. Parnachev and
B. Zwiebach for useful discussions.  The work of V.N. is supported by 
DOE grant DE-FG02-90ER40560.
The work of N.P. is partially supported by funds
provided by DOE under the cooperative research
agreement DE-FC02-94ER40818.

\

\

\

\

%%%%%%%%%%%%%%%%%%%%%%%%%%%%%%%%%%%%%%%%%%%%%%%%%%%%%%%%%%%%%%%%%%%%%%%%%
\break
%\vskip 1cm
\noindent{\bf Appendix A}

\

By definition we have
\eqn\dom{\eqalign{
& d\omega(\delta_i G, \delta_j G, \delta_k G) = (-)^{\epsilon_j}
\int \prod_{\beta=1}^3 d\tau_{\beta} d\theta_{\beta} \bigg (
\bigg \langle (b_{-1}\delta_i G) (Y \delta_j G) (Y \delta_k G) \bigg \rangle \cr
+& (-1)^{\epsilon_i} 
\bigg \langle (Y\delta_i G) (b_{-1} \delta_j G) (Y \delta_k G) \bigg \rangle
+(-1)^{\epsilon_i+\epsilon_j} 
\bigg \langle (Y\delta_i G) (Y \delta_j G) (b_{-1} \delta_k G) \bigg \rangle
\bigg ),
}}
with $\epsilon_i=\epsilon (\delta_i G)$.

Let us consider the following Ward identities for $b_{-1}$

\vskip .5cm

\centerline{$
\langle b_{-1} ( \delta_i G (Y\delta_jG) (Y\delta_k G) ) 
\rangle = 0 \Rightarrow $}
\centerline{$
\langle (b_{-1}\delta_i G) (Y\delta_j G) (Y\delta_k G)  \rangle +
(-)^{\epsilon_i} \langle \delta_i G (b_{-1}Y\delta_j G) (Y\delta_k G)  \rangle +
$}
\centerline{$
(-)^{\epsilon_i+\epsilon_j}
\langle \delta_i G (Y\delta_j) G (b_{-1}Y\delta_k G)  \rangle = 0,$}

\

\centerline{$
\langle b_{-1}  ( (Y\delta_i G) \delta_jG (Y\delta_k G) ) 
\rangle = 0 \Rightarrow $}
\centerline{$
\langle (b_{-1}Y\delta_i G)  \delta_j G (Y\delta_k G)  \rangle +
(-)^{\epsilon_i} \langle (Y\delta_i G) (b_{-1}\delta_j G) (Y\delta_k G)  \rangle +
$}
\centerline{$
(-)^{\epsilon_i+\epsilon_j}
\langle (Y\delta_i G) \delta_j G (b_{-1}Y\delta_k G)  \rangle = 0,$}

\

\centerline{$
\langle b_{-1} ( (Y\delta_i G) (Y\delta_jG) \delta_k G ) 
\rangle = 0 \Rightarrow $}
\centerline{$
\langle (b_{-1}Y\delta_i G)  (Y\delta_j G) \delta_k G  \rangle +
(-)^{\epsilon_i} \langle (Y\delta_i G) (b_{-1}Y\delta_j G) \delta_k G  \rangle +
$}
\centerline{$
(-)^{\epsilon_i+\epsilon_j}
\langle (Y\delta_i G) (Y\delta_j G) (b_{-1}\delta_k G)  \rangle = 0.$}

\vskip .5cm

Adding the last three identities, separating the appropriate terms 
and integrating gives

\centerline{$
d\omega(\delta_i G, \delta_j G, \delta_k G) =
$}
\centerline{$
(-)^{\epsilon_j+1} \int \prod_{\beta=1}^3 d\tau_{\beta} d\theta_{\beta} \bigg (
(-)^{\epsilon_i} \langle \delta_i G (b_{-1}Y\delta_j G) (Y\delta_k G)  \rangle +
(-)^{\epsilon_i+\epsilon_j}
\langle \delta_i G (Y\delta_j G) (b_{-1}Y\delta_k G)  \rangle + $}
\centerline{$
\langle (b_{-1}Y\delta_i G)  \delta_j G (Y\delta_k G)  \rangle +
(-)^{\epsilon_i+\epsilon_j}
\langle (Y\delta_i G) \delta_j G (b_{-1}Y\delta_k G)  \rangle + $}
\centerline{$
\langle (b_{-1}Y\delta_i G)  (Y\delta_j G) \delta_k G  \rangle +
(-)^{\epsilon_i} \langle (Y\delta_i G) (b_{-1}Y\delta_j G) \delta_k G  \rangle
\bigg ).
$}

\vskip .5cm

We have three pairs of terms, each of them labelled by the same statistics factor in front.
These pairs are actually vanishing. To see that, let us consider for example the pair
\eqn\pair{
\langle (b_{-1}Y\delta_i G)  \delta_j G (Y\delta_k G)  \rangle +
\langle (b_{-1}Y\delta_i G)  (Y\delta_j G) \delta_k G  \rangle.
}

\vskip .2cm

For the unperturbed correlator $\langle ... \rangle_0$
we can write the following two Ward identities for the BRST charge $Q$: 

\vskip .5cm

\centerline{$
\bigg \langle \xi (0) Q \bigg ( (b_{-1}Y\delta_i G) (\tau_1,\theta_1) 
(\xi Y\delta_j G)(\tau_2,\theta_2) (Y\delta_k G)(\tau_3,\theta_3) e^{\int d\tau
d\theta {\cal V}} \bigg ) \bigg \rangle_0 =
$}
\centerline{$
\bigg \langle X(0)(b_{-1}Y\delta_i G) (\tau_1,\theta_1) 
(\xi Y\delta_j G)(\tau_2,\theta_2) (Y\delta_k G)(\tau_3,\theta_3) \bigg \rangle
\Rightarrow 
$}
\centerline{$
\langle \xi (0) (Q b_{-1} Y\delta_i G) (\xi Y\delta_j G) (Y\delta_k G) \rangle +
$}
\centerline{$
(-)^{1+\epsilon_i}
\langle \xi(0) (b_{-1}Y\delta_i G) \delta_j G (Y\delta_k G) \rangle +
(-)^{\epsilon_i}
\langle \xi (0) (b_{-1}Y\delta_i G) (\xi (Q Y \delta_j G)) (Y\delta_k G) \rangle +
$}
\centerline{$
(-)^{\epsilon_i + \epsilon_j}
\langle \xi (0) (b_{-1}Y\delta_i G) (\xi Y\delta_j G) (Q Y\delta_k G) \rangle +
$}
\centerline{$
(-)^{\epsilon_i + \epsilon_j + \epsilon_k}
\langle \xi(0) (b_{-1}Y\delta_i G) (\xi Y\delta_j G) (Y\delta_k G)
[Q,e^{\int d\tau d\theta {\cal V}}]
\rangle_0 = 
$}
\centerline{$
\langle X(0)(b_{-1}Y\delta_i G) (\xi Y\delta_j G) (Y\delta_k G) \rangle
$}

\vskip .5cm

and

\vskip .5cm

\centerline{$
\bigg \langle \xi (0) Q \bigg ( (b_{-1}Y\delta_i G) (\tau_1,\theta_1)
(Y\delta_j G)(\tau_2,\theta_2) (\xi Y\delta_k G)(\tau_3,\theta_3) e^{\int d\tau
d\theta {\cal V}} \bigg ) \bigg \rangle_0 = 
$}
\centerline{$
\bigg \langle X(0)(b_{-1}Y\delta_i G) (\tau_1,\theta_1) 
(Y\delta_j G)(\tau_2,\theta_2) (\xi Y\delta_k G)(\tau_3,\theta_3) \bigg \rangle
\Rightarrow 
$}
\centerline{$ 
\langle \xi(0) (Q b_{-1} Y\delta_i G) (Y\delta_j G) (\xi Y\delta_k G) \rangle +
$}
\centerline{$
(-)^{1 + \epsilon_i}
\langle \xi(0) (b_{-1}Y\delta_i G) (Q Y \delta_j G) (\xi Y\delta_k G) \rangle +
(-)^{1+\epsilon_i + \epsilon_j}
\langle \xi(0) (b_{-1}Y\delta_i G) (Y\delta_j G) \delta_k G \rangle +$}
\centerline{$
(-)^{\epsilon_i + \epsilon_j}
\langle \xi(0) (b_{-1}Y\delta_i G)  (Y\delta_j G) (\xi (Q Y\delta_k G)) \rangle +
$}
\centerline{$
(-)^{\epsilon_i + \epsilon_j + \epsilon_k}
\langle \xi(0) (b_{-1}Y\delta_i G)  (Y\delta_j G) (\xi Y\delta_k G) 
[Q,e^{\int d\tau d\theta {\cal V}}]
\rangle_0 = 
$}
\centerline{$
\langle X(0)(b_{-1}Y\delta_i G) (Y\delta_j G) (\xi Y\delta_k G) \rangle.
$}

\vskip .5cm

We have explicitly inserted
a $\xi$ insertion in the center of the disk to saturate the zero mode of the $\xi \eta$
system \polch. This insertion is absent from similar expressions in the main text, but 
its presence is always implied.
The left hand side of these Ward identities is obtained by pushing the BRST current
contour onto the boundary and the right hand side by shrinking it to zero radius around the
center of the disk.

Solving in the above identities
for $\langle \xi (0) (b_{-1} Y\delta_i G) \delta_j G (Y \delta_k G) \rangle$ and
$\langle \xi (0) (b_{-1} Y \delta_i G) (Y\delta_j G) \delta_k G \rangle$ and adding the 
resulting expressions gives

\vskip .5cm

\centerline{$
\langle \xi (0) (b_{-1} Y\delta_i G) \delta_j G (Y \delta_k G) \rangle+
\langle \xi (0) (b_{-1} Y\delta_i G) (Y\delta_j G) \delta_k G \rangle=
$}
\centerline{$ 
(-)^{\epsilon_i}
\langle \xi (0) (Q b_{-1} Y\delta_i G) (\xi Y\delta_j G)  (Y\delta_k G) \rangle +
(-)^{\epsilon_i+\epsilon_j}
\langle \xi(0) (Q b_{-1} Y\delta_i G) (Y\delta_j G) (\xi Y\delta_k G) \rangle +
$}
\centerline{$
\langle \xi(0) (b_{-1}Y\delta_i G) (\xi (Q Y \delta_j G))  (Y\delta_k G) \rangle +
(-)^{\epsilon_j + 1}
\langle \xi(0) (b_{-1}Y\delta_i G) (Q Y \delta_j G) (\xi Y\delta_k G) \rangle +
$}
\centerline{$
(-)^{\epsilon_i+1}
\langle X(0)(b_{-1}Y\delta_i G) (\xi Y\delta_j G) (Y\delta_k G) \rangle+
$}
\centerline{$
(-)^{\epsilon_j}
\langle \xi(0) (b_{-1}Y\delta_i G) (\xi  Y\delta_j G) (Q Y\delta_k G) \rangle +
\langle \xi(0)(b_{-1}Y\delta_i G) (Y\delta_j G) (\xi (Q Y\delta_k G)) \rangle +
$}
\centerline{$
(-)^{\epsilon_j+\epsilon_k}
\langle \xi(0) (b_{-1}Y\delta_i G) (\xi  Y\delta_j G)  (Y\delta_k G) [Q, e^{\int d\tau
d\theta {\cal V}}]
\rangle_0 +
$}
\centerline{$
(-)^{\epsilon_k}
\langle \xi(0) (b_{-1}Y\delta_i G)  (Y\delta_j G) (\xi Y\delta_k G) [Q,e^{\int d\tau
d\theta {\cal V}}]
\rangle_0 +
$}
\centerline{$
(-)^{\epsilon_j+\epsilon_i+1}
\langle X(0)(b_{-1}Y\delta_i G) (Y\delta_j  G) (\xi Y\delta_k G) \rangle.
$}

\vskip .5cm

Since, the position of the $\xi$ insertions is irrelevant, we can
move appropriately the $\xi$ insertion (on the boundary) 
in the above expressions and after integrating 
over $\tau$ and $\theta$ we take

\eqn\pairfinal{
\int \prod_{\beta=1}^3 d\tau_{\beta} d\theta_{\beta} \bigg (
\langle (b_{-1}Y\delta_i G)  \delta_j G Y\delta_k G  \rangle +
\langle (b_{-1}Y\delta_i G)  Y\delta_j G \delta_k G  \rangle
\bigg )=0.
}

\vskip .5cm

Continuing in the same fashion for the other two pairs we conclude that
$d\omega=0$.

\vskip .5cm

The proof of $V$-invariance of $\omega$ goes in a similar way. The only
addition is the use of the identity  
$\{Q,b_{-1} \}=v^a \partial_a$, where $v^a \partial_a$ is the generator
of rotations on the disk, as well as the use of the identity
\eqn\totalder{
\int d\tau d\theta v^a \partial_a (Y \delta G)=0 .
}

%%%%%%%%%%%%%%%%%%%%%%%%%%%%%%%%%%%%%%%%%%%%%%%%%%%%%%%%%%%%%%%%%%%%%%%%%

\vskip 0.5cm
\noindent{\bf Appendix B}

\

In this appendix we would like to show equation \yqcv. From \qcv \ we have
\eqn\qcvv{
[Q,C{\cal V}]=[(1-h)C\partial_{\tau}C-\frac{1}{4}(D_{\theta}C)(D_{\theta}C)]{\cal V}+
\frac{1}{2}C(D_{\theta}C)(D_{\theta}{\cal V}).
}
In components, the above equation involves the expressions 
\foot{All products of fields appearing here are normal
ordered in the usual CFT fashion, i.e. 
$ AB(w) = \oint \frac{dz}{2\pi i} \frac{1}{z-w}A(z)B(w) $.}
\eqn\cdc{
C\partial_{\tau}C=c\partial_{\tau}c+\theta(\gamma \partial_{\tau}c-c\partial_{\tau}
\gamma ),
}
\eqn\dcdc{
(D_{\theta}C)(D_{\theta}C)=\gamma^2 + 2\theta \gamma \partial_{\tau} c,
}
\eqn\cDc{
CD_{\theta}C=c\gamma+\theta (\gamma^2 - c\partial_{\tau}c).
}
Since $Y=-\partial \xi c e^{-2\phi}$ we easily deduce by using the relevant 
OPEs that
\eqn\ygamma{
Y\gamma=-ce^{-\phi}.
}
Hence, acting with $Y$ on the above equations gives the following expressions
\eqn\ycdc{
YC\partial_{\tau}C=Yc \partial_{\tau}c+
\theta Y\gamma\partial_{\tau}c=Yc \partial_{\tau} c
-\theta c\partial_{\tau}ce^{-\phi},
}
\eqn\ydcdc{
Y(D_{\theta}C)(D_{\theta}C)=Y\gamma^2+2\theta Y\gamma\partial_{\tau}c=
Y\gamma^2-2\theta c\partial_{\tau}c 
e^{-\phi},
}
\eqn\ycDc{
YCD_{\theta}C=\theta(Y\gamma^2-Yc\partial_{\tau}c).
}
Combining these equations with \qcvv \ gives

\centerline{$
Y[Q,C{\cal V}]=\bigg ( (1-h)(Yc\partial_{\tau}c-
\theta c\partial_{\tau}ce^{-\phi}-\frac{1}{4} Y\gamma^2
+2\frac{1}{4}\theta
c\partial_{\tau}ce^{-\phi}\bigg ) {\cal V}+
\frac{1}{2} \theta (Y\gamma^2-Yc\partial_{\tau}c)D_{\theta}{\cal V}=
$}
\centerline{$
\bigg ( (1-h) Yc\partial_{\tau}c-\frac{1}{4}Y\gamma^2 \bigg ) {\cal V}+
\frac{1}{2}\theta (Y\gamma^2-Yc\partial_{\tau}c)D_{\theta} {\cal V}+
\bigg( h-\frac{1}{2}\bigg) \theta c\partial_{\tau}ce^{-\phi}{\cal V}=
$}
\eqn\yyqcv{
\bigg ( (1-h) Yc\partial_{\tau}c-\frac{1}{4}Y\gamma^2 \bigg ) {\cal V}+
\frac{1}{2}\theta (Y\gamma^2-Yc\partial_{\tau}c)D_{\theta} {\cal V}+
\bigg( h-\frac{1}{2}\bigg) \theta c\partial_{\tau}ce^{-\phi}D,
}
i.e. equation \yqcv .

\vskip 0.5cm
\noindent{\bf Appendix C}

\

In this appendix we use the SUSY Ward identities on the real line
in order to demonstrate relations between the coefficients of certain 2-point
and 3-point functions at the conformal point.

\vskip .5cm
{\it 2-point functions}

\medskip

The SUSY Ward identity for the 2-point functions on the real line reads
\eqn\wardsusytwo{\eqalign{&
\delta \langle U_i (x_1) D_j (x_2) \rangle_0 =0 \Rightarrow
\cr 
\langle \partial_{x_1} D_i(x_1) & D_j(x_2) \rangle_0 +
\langle U_i(x_1) U_j(x_2) \rangle_0 = 0,
}}
where $\delta$ denotes an infinitesimal SUSY transformation. 

Plugging in \wardsusytwo \ the CFT expressions
\eqn\dd{
\langle D_i(x_1) D_j(x_2) \rangle_0 = 
\frac{c \delta_{h_i,h_j}}{ |x_1 - x_2 |^{2 h}}  {\rm sign}(x_1-x_2)
}
and
\eqn\uu{
\langle U_i(x_1) U_j(x_2) \rangle_0 = 
\frac{c \delta_{h_i,h_j}}{ |x_1 - x_2 |^{2 h + 1}} 
}
with $h=h_i=h_j$, we conclude that 
\eqn\cbbbb{
2 c h = b .
}

\vskip .5cm
{\it 3-point functions}

\medskip

The relevant Ward identity  is
\eqn\wardsusythree{\eqalign{&
\delta \langle U_i (x_1) D_j(x_2) U_k(x_3) \rangle_0 \Rightarrow
\cr 
\langle \partial_{x_1} D_i(x_1) D_j(x_2) U_k(x_3) \rangle_0  +
\langle & U_i(x_1) U_j(x_2) U_k(x_3) \rangle_0 -
\langle U_i(x_1) D_j(x_2) \partial_{x_3} D_k(x_3)  \rangle_0 = 0.
}}
Substituting for the CFT expressions 
\eqn\uuu{
\langle U_i(x_1) U_j(x_2) U_k(x_3)  \rangle_0 =
\frac{B_{ijk}} 
{|x_1-x_2 |^{h_i+h_j-h_k+\frac{1}{2}}
 |x_2-x_3 |^{h_k+h_j-h_i+\frac{1}{2}}
 |x_3-x_1 |^{h_i+h_k-h_j+\frac{1}{2}} },
}
\eqn\ddu{
\langle D_i (x_1) D_j (x_2) U_k(x_3) \rangle_0 = 
\frac{C_{ijk} \ \ {\rm sign}(x_1-x_2)}{ |x_1-x_2 |^{
h_i+h_j-h_k-\frac{1}{2}}
 |x_2-x_3  |^{h_j+h_k-h_i+\frac{1}{2}}
 |x_3-x_1  |^{h_i+h_k-h_j+\frac{1}{2}} }
}
gives  
\eqn\bijkcijkone{\eqalign{
B_{ijk}(x_1 &-x_3) = C_{ijk}\bigg(h_i+h_j-h_k-\frac{1}{2}\bigg)(x_1-x_3) +
C_{ijk}\bigg(h_k+h_i-h_j+\frac{1}{2}\bigg) (x_1-x_2) + 
\cr &
+ C_{jki}\bigg(h_j+h_k-h_i-\frac{1}{2}\bigg) (x_1-x_3)
- C_{jki}\bigg (h_i+h_k-h_j+\frac{1}{2} \bigg) (x_3-x_2).
}}
This equation must be valid for any value of the worldsheet variables $x_1,x_2,x_3$. Hence,
\eqn\bijkcijktwo{
B_{ijk}=2h_i C_{ijk}+C_{jki}\bigg (h_k+h_j-h_i-\frac{1}{2}\bigg),
}
\eqn\bijkcijkthree{
B_{ijk}=C_{ijk}(h_i+h_j-h_k)+2h_k C_{jki},
}
\eqn\bijkcijkfour{
C_{ijk}=C_{jki}.
}
Equivalently, the first two equations give
\eqn\bccon{
B_{ijk}=-C_{ijk}\bigg (\frac{1}{2}-h_i-h_j-h_k \bigg).
}

\listrefs
\end